\documentclass[prd,superscriptaddress,nofootinbib,amsmath,amssymb,aps,11pt]{revtex4}

\usepackage{bm}
\usepackage{amsfonts}
\usepackage{latexsym}
\usepackage[latin1]{inputenc}
\usepackage{graphicx}
\usepackage{amsmath}
\usepackage{palatino}
\usepackage{mathpazo}
\usepackage{booktabs}
\usepackage{dcolumn}

\def\jnl@style{\it}
\def\aaref@jnl#1{{\jnl@style#1}}

\def\aaref@jnl#1{{\jnl@style#1}}

\def\aj{\aaref@jnl{AJ}}                   
\def\apj{\aaref@jnl{ApJ}}                 
\def\apjl{\aaref@jnl{ApJ}}                
\def\apjs{\aaref@jnl{ApJS}}               
\def\apss{\aaref@jnl{Ap\&SS}}             
\def\aap{\aaref@jnl{A\&A}}                
\def\aapr{\aaref@jnl{A\&A~Rev.}}          
\def\aaps{\aaref@jnl{A\&AS}}              
\def\mnras{\aaref@jnl{Mon.~Not.~Roy.~Astron.~Soc.}}             
\def\prd{\aaref@jnl{Phys.~Rev.~D}}        
\def\prc{\aaref@jnl{Phys.~Rev.~C}}  
\def\prl{\aaref@jnl{Phys.~Rev.~Lett.}}    
\def\qjras{\aaref@jnl{QJRAS}}             
\def\skytel{\aaref@jnl{S\&T}}             
\def\ssr{\aaref@jnl{Space~Sci.~Rev.}}     
\def\zap{\aaref@jnl{ZAp}}                 
\def\nat{\aaref@jnl{Nature}}              
\def\aplett{\aaref@jnl{Astrophys.~Lett.}} 
\def\apspr{\aaref@jnl{Astrophys.~Space~Phys.~Res.}} 
\def\physrep{\aaref@jnl{Phys.~Rep.}}      
\def\physscr{\aaref@jnl{Phys.~Scr}}       
\def\commat{\aaref@jnl{Comm.~Math.~Phys.}}              
\def\science{\aaref@jnl{Science}}               
\def\cqg{\aaref@jnl{Classical Quant.~Grav.}}            
\def\jpcs{\aaref@jnl{JPCS}}                                     
\def\ijmpd{\aaref@jnl{Int.~J.~Mod.~Phys.~D}}                    
\def\grg{\aaref@jnl{Gen.~Relat.~Gravit.}}               
\def\rpp{\aaref@jnl{Rep.~Prog.~Phys.}}          
\def\npa{\aaref@jnl{Nucl.~Phys.~A}}        
\def\lrr{\aaref@jnl{Living Rev.~Rel.}}                   
\def\jcap{\aaref@jnl{J.~Cosmology Astropart.~Phys.}}    
\def\rmp{\aaref@jnl{Rev.~Mod.~Phys.}}   

\allowdisplaybreaks[1]

\begin{document}

\title{No-hair theorems for non-canonical self-gravitating static  multiple scalar fields}

\author{Daniela D. Doneva}
\email{daniela.doneva@uni-tuebingen.de}
\affiliation{Theoretical Astrophysics, Eberhard Karls University of T\"ubingen, T\"ubingen 72076, Germany}
\affiliation{INRNE - Bulgarian Academy of Sciences, 1784  Sofia, Bulgaria}

\author{Stoytcho S. Yazadjiev}
\email{yazad@phys.uni-sofia.bg}
\affiliation{Theoretical Astrophysics, Eberhard Karls University of T\"ubingen, T\"ubingen 72076, Germany}
\affiliation{Department of Theoretical Physics, Faculty of Physics, Sofia University, Sofia 1164, Bulgaria}
\affiliation{Institute of Mathematics and Informatics, 	Bulgarian Academy of Sciences, 	Acad. G. Bonchev St. 8, Sofia 1113, Bulgaria}


\begin{abstract}
We prove under certain assumptions no-hair theorems for non-canonical self-gravitating static  multiple scalar fields in spherically symmetric spacetimes.  It is shown that the only static, spherically symmetric and asymptotically flat  black hole solutions consist of  the Schwarzschild metric and a constant multi-scalar map. We also prove that there are  no static, horizonless, asymptotically flat,  spherically symmetric solutions with static scalar fields and a regular center. The last theorem shows that  the static, asymptotically flat,  spherically symmetric reflecting compact objects with Neumann boundary conditions  can not support a  non-trivial   self-gravitating non-canonical (and canonical) multi-scalar map  in their exterior spacetime regions. In order to prove the no-hair theorems we derive a new divergence identity. 
\end{abstract}

\maketitle

\section{Introduction}

Many theories beyond the standard model physics predict the existence of additional scalar fields. The natural question that arises is whether
there exist black holes supporting scalar hair beyond those existing in General relativity. This is an old problem in gravitational physics going back to the classical paper \cite{Bekenstein_1972}. In the case of a single scalar field, minimally or non-minimally coupled to gravity,  with a non-negative potential, there are no-hair theorems which rule out the existence of stationary asymptotically flat black holes with a scalar hair under some certain conditions. We refer the reader to the recent review \cite{Herdeiro_2015} for exhaustive discussion of the black holes with single scalar hair. Another interesting and important problem is the extension of the no-hair theorems to the case of a single non-canonical scalar field. No-hair theorems for a non-canonical single scalar field were   presented in \cite{Graham_2014a}-\cite{Zhomga_2016}.

It is of obvious theoretical interest to see how far the above results can be extended in the case of both canonical and non-canonical multiple scalar fields. Some results in this direction have also been established for canonical multiple scalar fields. It was proven in \cite{Heusler_1992} that there exist no static and spherically symmetric  asymptotically flat black holes with multiple scalar hair when the scalar fields  are static and their potential is non-negative. Similar result holds for   static, horizonless, asymptotically flat,  spherically symmetric solutions with static scalar fields \cite{Heusler_1992}. These results can be extended to the case of stationary (rotating) solutions with time independent harmonic  mappings \cite{Heusler_1995}.  In general the situation with the  multiple scalar fields is much more complicated in comparison with the case of a single scalar field. For example, if we allow  the scalar fields to be time dependent in a certain way so that the effective scalar fields energy-momentum tensor to be time independent, the picture can change drastically.
It was shown in \cite{Herdeiro_2014} that the Kerr black hole can support hair of two time dependent scalar fields. Generalizations of \cite{Herdeiro_2014} were presented in \cite{Collodel_2020a}. Moreover there exist also soliton-like (horizoless) solutions with multiple scalar fields \cite{Yazadjiev_2019}, \cite{Collodel_2020b}.     

The purpose of the present paper is to prove several no-hair theorems for  non-canonical self-gravitating static  multiple scalar fields
in spherically symmetric spacetimes. Under certain assumptions we prove that there are no non-trivial static, asymptotically flat and spherically symmetric black holes and particle-like solutions with static non-canonical multiple scalar hair. We also show that the static, spherically symmetric  reflecting objects with Neumann boundary conditions in an asymptotically flat spacetime  can not support static hair consisting of non-canonical scalar fields. We prove our no-hair theorems by first deriving a new divergence identity. In the particular case of canonical scalar fields our approach gives a new independent proof of the theorems of \cite{Heusler_1992}.

\section{Basic equations}

Let $(M, g)$ is the 4-dimensional spacetime and $({\cal E}_N,\gamma)$ is a $N$-dimensional Riemannian manifold with metric $\gamma$. We consider a map $\varphi: (M,g)\to ({\cal E}_N,\gamma)$  and its deferential $d\varphi$ induces a map between the tangent spaces of $M$ and ${\cal E}_N$, $d\varphi: TM\to T{\cal E}_N$. The norm of the differential will be denoted by $<d\varphi,d\varphi>$. In local coordinate patches on 
$M$ and ${\cal E}_N$ we have 

\begin{eqnarray}
<d\varphi,d\varphi>= g^{\mu\nu}(x)\gamma_{ab}(\varphi(x))\partial_\mu\varphi^a(x)\partial_\nu\varphi^b(x).
\end{eqnarray} 

The multi-scalar theories we focus on are self-gravitating theories with non-canonical scalar fields given by the action

\begin{eqnarray}\label{action}
S=\frac{1}{16\pi G} \int_{M} d^4x\sqrt{-g}\left[R - f_{*}(\varphi,K)\right],
\end{eqnarray}
 where  $R$ is the Ricci scalar curvature,  $K=2<d\varphi,d\varphi>$ and  $f_{*}(\varphi,K)$ is at least  $C^2$ - function in $\varphi$ and $K$. Some mild restrictions on $f_{*}(\varphi,K)$
 will be discussed  below. The canonical scalar fields correspond to the particular choice $f_{*}(\varphi,K)= K + 4V(\varphi)$, where $V(\varphi)$ is the scalar fields potential. The field equations corresponding to the action (\ref{action}) and written in local coordinate patches of  $M$ and ${\cal E}_N$ are the following 
 
 \begin{eqnarray}\label{FE1}
 R_{\mu\nu} - \frac{1}{2}R g_{\mu\nu} = 2\frac{\partial f_*}{\partial K}\gamma_{ab}(\varphi)\nabla_{\mu}\varphi^a \nabla_{\nu}\varphi^b - 
 \frac{1}{2} f_{*} g_{\mu\nu}, \\
\nabla_{\mu}\left(\frac{\partial f_*}{\partial K}\nabla^{\mu}\varphi^a\right)= 
-\frac{\partial f_*}{\partial K} \gamma^{a}_{cd}(\varphi)\nabla_{\mu}\varphi^c \nabla^{\mu}\varphi^d + \frac{1}{4}\gamma^{ab}(\varphi)\frac{\partial f_{*}}{\partial\varphi^b}, \label{FE2}
 \end{eqnarray}
where $\nabla_{\mu}$ is the covariant derivative with respect to the spacetime metric $g_{\mu\nu}$ and 
$\gamma^{a}_{cd}(\varphi)$ are the Christoffel symbols with respect to the target space metric $\gamma_{ab}(\varphi)$.

Let us discuss the restrictions on the function $f_*(\varphi, K)$. As in the case of a single non-canonical scalar field, 
the classical and quantum stability of the theory  is guaranteed when $ \frac{\partial f_*}{\partial K}\ge 0$ and 
$\frac{\partial f_*}{\partial K} + 2K \frac{\partial^2 f_*}{\partial^2 K}\ge 0$ \cite{Picon_2005}. These conditions ensure also that the  initial value formulation of the theory is well-posed \cite{Picon_2005}. Another condition which is natural from a physical point of view is 
$\lim_{K\to 0} \frac{\partial f_*}{\partial K}=1$ which means that that in the weak field regime we recover the canonical case.
One more condition on $f_*$  playing crucial role in proving the no-hair theorems,  will be imposed in Sec. 3.

From now on we shall focus on the static and spherically symmetric case with a metric 
\begin{eqnarray}
ds^2= - e^{2\Phi(r)}dt^2 + e^{2\Lambda(r)}dr^2 + r^2 s_{ij}dx^idx^j,
\end{eqnarray}  
where $s_{ij}$ is the metric on the unit 2D sphere, namely $s_{ij}dx^idx^j=d\theta^2  + \sin^2\theta d\phi^2$.

We require the scalar fields to be also static, ${\cal L}_{\xi}\varphi^a=0$, i.e. the Lie derivative of the scalar fields along the timelike
Killing vector $\xi=\frac{\partial}{\partial t}$ to be zero. This condition automatically insures that the effective energy-momentum tensor 
of the scalar fields is static, ${\cal L}_{\xi} T^{\varphi}_{\mu\nu}=0$. 
Contrary to the static symmetry, the scalar fields are not required to respect the spherical symmetry, i.e. to depend only on the radial coordinate $r$. They can depend on the angular  variables $\theta$ and $\phi$ in such way that the energy-momentum tensor respects the spherical symmetry (see for example \cite{Doneva_2020a,Doneva_2020b}.

Under our assumptions the dimensionaly reduced field equations are the following 
\begin{eqnarray}
	&&\frac{2}{r}e^{-2\Lambda} \Lambda^{\prime} + \frac{1}{r^2}\left(1-e^{-2\Lambda}\right)= \frac{1}{2}f_*, \label{DRE} \\ 
&&\frac{2}{r}e^{-2\Lambda} \Phi^{\prime} - \frac{1}{r^2}\left(1-e^{-2\Lambda}\right)= 
2 \frac{\partial f_*}{\partial K}e^{-2\Lambda}\gamma_{ab}(\varphi)\partial_r\varphi^a \partial_r\varphi^b - \frac{1}{2}f_*, \\\
&&e^{-2\Lambda}\left[\Phi^{\prime\prime} + (\Phi^{\prime} + \frac{1}{r})(\Phi^{\prime} - \Lambda^{\prime})\right]r^{2} s_{ij}=
2 \frac{\partial f_*}{\partial K}\gamma_{ab}(\varphi)\partial_i\varphi^a \partial_j\varphi^b -  \frac{1}{2}f_* r^2 s_{ij}, \\
&&\partial_r \left(e^{\Phi-\Lambda} r^2 \frac{\partial f_*}{\partial K} \gamma_{ab}(\varphi)\partial_r\varphi^b \right) +
\frac{e^{\Phi + \Lambda}}{\sqrt{s}} \partial_i\left(\sqrt{s} s^{ij}\frac{\partial f_*}{\partial K} \gamma_{ab}(\varphi)\partial_j\varphi^b\right) = \frac{e^{\Phi + \Lambda}}{4} r^2 \left(\frac{\partial f_*}{\partial \varphi^a} + 
\frac{\partial f_*}{\partial K} \frac{\partial K}{\partial \varphi^a}\right).\label{EQFF}
\end{eqnarray} 

From the spherical symmetry via the dimensionally reduced field equations it follows that $f_*$,  $\frac{\partial f_*}{\partial K}\gamma_{ab}(\varphi)\partial_r\varphi^a \partial_r\varphi^b $ and $\frac{\partial f_*}{\partial K}s^{ij}\gamma_{ab}(\varphi)\partial_i\varphi^a \partial_j\varphi^b$ are functions of $r$ only. We define
 
\begin{eqnarray}
P^2(r)=\frac{\partial f_*}{\partial K}\gamma_{ab}(\varphi)\partial_r\varphi^a \partial_r\varphi^b, \,\,
H^{2}(r)= \frac{\partial f_*}{\partial K}s^{ij}\gamma_{ab}(\varphi)\partial_i\varphi^a \partial_j\varphi^b .
\end{eqnarray}

In the present paper we consider only asymptotically flat spacetimes. In this case, by using the dimensionally 
reduced field equations, it is not difficult one to show  that $\lim_{r\to \infty} f_{*}(r)=0$, $\lim_{r\to \infty} P^2(r)=0$ and $\lim_{r\to \infty} H^2(r)=0$. More precisely   $f_{*}(r)$, $P^2(r)$ and  $H^2(r)$  drop off  at least  as  

\begin{eqnarray}
f_*(r) \sim \frac{1}{r^4},  \,\,\, P^2(r) \sim \frac{1}{r^4} , \,\,\, H^2(r)\sim \frac{1}{r^2}
\end{eqnarray}
for $r\to \infty$.  

\section{Divergence identity}

In this section we derive a divergence identity which plays a central role in proving the no-hair theorems. 
In fact the potential application of  our divergence identity is beyond the no-hair theorems. It can be used to study 
the quantitative and qualitative properties of static and spherically symmetric vacuum solutions in theories with multiple scalar fields
as those presented in \cite{Doneva_2020a}-\cite{Zhdanov_2020}.
 
 In order to derive the desired divergence identity we multiply the equation (\ref{EQFF}) for $\varphi^a$ by $\partial_r\varphi^a$ and after some algebra we obtain 

\begin{eqnarray}\label{I1}
&&\partial_r\left(e^{\Phi-\Lambda} r^2 P^2(r)\right) 
- e^{\Phi-\Lambda} r^2 \frac{\partial f_*}{\partial K}\gamma_{ab}(\varphi)\partial_r\varphi^b \partial^2_r\varphi^a + 
 e^{\Phi + \Lambda} \partial_r\varphi^a {\cal D}_i J^i_a \nonumber \\ 
 &&=\frac{e^{\Phi + \Lambda}}{4} r^2 \left(\frac{\partial f_*}{\partial \varphi^a} + 
 \frac{\partial f_*}{\partial K} \frac{\partial K}{\partial \varphi^a}\right) \partial_r\varphi^a 
\end{eqnarray}
where we have defined $J^i_a=\frac{\partial f_*}{\partial K}\gamma_{ab}(\varphi){\cal D}^i\varphi^b$ and ${\cal D}_i$ is the covariant derivative with respect to $s_{ij}$. The next step is to express $e^{\Phi-\Lambda} r^2 \frac{\partial f_*}{\partial K} \partial_r^2 \varphi^a$
again from (\ref{EQFF}) and to substitute into the above equation (\ref{I1}). We find

\begin{eqnarray}\label{I2}
&&\partial_r\left(e^{\Phi-\Lambda} r^2 P^2(r)\right) 
+ \partial_r \left(e^{\Phi-\Lambda} r^2 \frac{\partial f_*}{\partial K}\gamma_{ab}(\varphi)\right)\partial_r\varphi^a\partial_r\varphi^b
+ 2 e^{\Phi  + \Lambda} \partial_r\varphi^a {\cal D}_i J^i_a 
 \nonumber \\ 
&&=\frac{e^{\Phi + \Lambda}}{2} r^2 \left(\frac{\partial f_*}{\partial \varphi^a} + 
\frac{\partial f_*}{\partial K} \frac{\partial K}{\partial \varphi^a}\right) \partial_r\varphi^a .
\end{eqnarray}

Proceeding further we can work out the term $2 e^{\Phi  + \Lambda} \partial_r\varphi^a {\cal D}_i J^i_a$. Using the dimensional reduced field equations and after long manipulations one can show that 

\begin{eqnarray}\label{I2a}
2 e^{\Phi  + \Lambda} \partial_r\varphi^a {\cal D}_i J^i_a= - e^{\Phi+\Lambda} \partial_r H^2(r) + 
e^{\Phi+\Lambda}  {\cal D}_i\varphi^a {\cal D}^i\varphi^b \,\partial_r\left(\frac{\partial f_*}{\partial K}\gamma_{ab}(\varphi)\right).
\end{eqnarray}

Substituting this expression back into (\ref{I2}) and after long algebra we get 

\begin{eqnarray}\label{I3}
&&\frac{d}{dr}\left[e^{\Phi+\Lambda} H^2(r) + \frac{1}{2}e^{\Phi+\Lambda}r^2 (f_* - K \frac{\partial f_*}{\partial K}) -
e^{\Phi-\Lambda} r^2 P^2(r) \right] \\
&&=e^{\Phi-\Lambda}r^2 (\Phi^{\prime} -\Lambda^{\prime} + \frac{2}{r}) P^2(r) + 
e^{\Phi+\Lambda}(\Phi^{\prime} + \Lambda^{\prime}) \left[H^2(r) + \frac{1}{2}(f_* - K \frac{\partial f_*}{\partial K})r^2\right]
+ re^{\Phi+\Lambda} (f_* - K \frac{\partial f_*}{\partial K}) .\nonumber 
\end{eqnarray}

The last step is to express  $(\Phi^{\prime} + \Lambda^{\prime})$ and $(\Phi^{\prime} -\Lambda^{\prime} + \frac{2}{r})$ form 
the dimensionally reduced equations, namely 

\begin{eqnarray}\label{I3a}
&&\Phi^{\prime} + \Lambda^{\prime}= r P^2(r), \\
&&(\Phi^{\prime} -\Lambda^{\prime} + \frac{2}{r})e^{-2\Lambda}= \frac{1+ e^{-2\Lambda}}{r} + r e^{-2\Lambda}P^2(r) - \frac{1}{2}r f_* 
\end{eqnarray}
and to substitute them  back into (\ref{I3}). Doing so we finally obtain the desired divergence identity

\begin{eqnarray}\label{DI}
&&\frac{d}{dr}\left[e^{\Phi+\Lambda} H^2(r) + \frac{1}{2}e^{\Phi+\Lambda}r^2 (f_* - K \frac{\partial f_*}{\partial K}) -
e^{\Phi-\Lambda} r^2 P^2(r) \right] \\
&&= r e^{\Phi + \Lambda} \left[(1+ e^{-2\Lambda}) P^2(r) + (f_* - K \frac{\partial f_*}{\partial K}) \right]. \nonumber
\end{eqnarray}

In what follows we shall consider theories with 
\begin{eqnarray}\label{CON}
f_* - K \frac{\partial f_*}{\partial K}\ge 0.
\end{eqnarray}

When this condition  is satisfied, the right hand side of the divergence identity is non-negative. In the case of canonical scalar fields, 
i.e. for $f_*= K + 4V(\varphi)$, our condition (\ref{CON}) reduces just to $V(\varphi)\ge 0$. It is also instructive to present the divergence identity in the particular case of canonical scalar fields. Then the divergent identity takes the form 

\begin{eqnarray}\label{DICAN}
&&\frac{d}{dr}\left[e^{\Phi+\Lambda} H^2(r) + 2e^{\Phi+\Lambda}r^2 V(\varphi) - 
e^{\Phi-\Lambda} r^2 P^2(r) \right] \\
&&= r e^{\Phi + \Lambda} \left[(1+ e^{-2\Lambda}) P^2(r) + 4V(\varphi) \right]. \nonumber
\end{eqnarray}

\section{No-hair theorems} 

In this section we present and prove our no-hair theorems. The first no-hair theorem is a theorem which states that there do not exist static, asymptotically flat  and spherically symmetric black holes with non-trivial non-canonical static multiple scalar fields. More precisely we have:

\medskip
\noindent

{\bf Theorem 1} {\it Let us  consider self-gravitating non-canonical multi-scalar map $\varphi: (M,g)\to ({\cal E}_N, \gamma)$ with the action 
(\ref{action}) and let the function $f_{*}(\varphi, K)$ satisfy the inequality $f_* - K \frac{\partial f_*}{\partial K}\ge 0$. 
Then every static and spherically symmetric black hole solution to the field equations (\ref{FE1})-(\ref{FE2}) with static 
multi-scalar map $\varphi$ (${\cal L}_\xi\varphi$=0)  and regular horizon   consists of the Schwarzschild solution  and a constant map $\varphi_0$  with $f_*(\varphi_0, K=0)=0$. } 
  
\medskip
\noindent

The key role in proving the theorem plays the divergence identity (\ref{DI}) derived in the previous section. Integrating this identity from the horizon $r=r_h$ to infinity we get 

\begin{eqnarray}\label{BHInt}
- e^{(\Phi+\Lambda)_h} \left[H^2(r_h) + \frac{1}{2}r_h^2 (f_* - K \frac{\partial f_*}{\partial K})_h\right]
= \int^{\infty}_{r_h} dr r e^{\Phi + \Lambda} \left[(1+ e^{-2\Lambda}) P^2(r) + (f_* - K \frac{\partial f_*}{\partial K}) \right], 
\end{eqnarray}
where we have taken into account that $\lim_{r\to \infty} H^2(r)=0$, $\lim_{r\to \infty} r^2 P^2(r)=0$, $\lim_{r\to \infty} r^2(f_* - K \frac{\partial f_*}{\partial K})=0$ as well as $(e^{\Phi-\Lambda} r^2 P^2(r))_h=0$. Note that $(\Phi+\Lambda)_h$ is finite for regular horizons.
The right hand side of (\ref{BHInt}) is non-negative while the left hand side is non-positive. Therefore we can conclude that
both sides vanish. Consequently we have $P^2(r)=0$ and $f_* - K \frac{\partial f_*}{\partial K}=0$ for every $r\in [r_h,\infty)$ as well as 
$H^2(r_h)=0$.  Turning again to the divergence identity (\ref{DI}) and using the fact that $P^2(r)=0$ and $f_* - K \frac{\partial f_*}{\partial K}=0$ we find that $H^2(r)=0$ for every $r\in [r_h,\infty)$. The fact that $P^2(r)=H^2(r)=0$ implies $K(r)=0$ which in turn means $f_*(\varphi,K=0)=0$. As a consequence of all this we  conclude that the map $\varphi$ is a constant map $\varphi=\varphi_0$ with $f_*(\varphi_0,K=0)=0$. 

All the above results show that the right hand side of the dimensionally reduced field equations (\ref{DRE})-(\ref{EQFF}) vanish and the equations are reduced to the vacuum, static and spherically symmetric Einstein equations whose unique black hole solution with regular horizon is the Schwarzshild one. This concludes the proof of the theorem. 

The next theorem deals with  static, horizonless, asymptotically flat,  spherically symmetric solutions with static scalar fields and a regular center. Before formulating the theorem let us discuss the regularity conditions at the center. The geometry of the $t=const$ hypersurfaces has to be  locally Euclidean in a small vicinity of the center $r=0$ which means $\lim_{r\to 0} e^{2\Lambda}=1$.      

\medskip
\noindent

{\bf Theorem 2} {\it Let us  consider self-gravitating non-canonical multi-scalar map $\varphi: (M,g)\to ({\cal E}_N, \gamma)$ with the action 
	(\ref{action}) and let the function $f_{*}(\varphi, K)$ satisfy the inequality $f_* - K \frac{\partial f_*}{\partial K}\ge 0$. 
	Then every static, asymptotically flat,  spherically symmetric solution 
	to the field equations (\ref{FE1}-\ref{FE2})  with static scalar fields and a regular center  consists of the flat metric   and a constant map $\varphi_0$  with $f_*(\varphi_0, K=0)=0$. } 

\medskip
\noindent

The proof is as follows. The regularity at the center and the dimensionally reduced field equations imply $\lim_{r\to 0} e^{\Phi+\Lambda}H^2(r)=
\lim_{r\to 0} r^2 e^{\Phi-\Lambda}P^2(r)=\lim_{r\to 0} r^2 e^{\Phi+ \Lambda} \left( f* - K \frac{\partial f_*}{\partial K}\right)=0$.  
Using this fact and integrating the divergence identity from $r=0$ to infinity we obtain 

\begin{eqnarray}
\int^{\infty}_{0} dr r e^{\Phi + \Lambda} \left[(1+ e^{-2\Lambda}) P^2(r) + (f_* - K \frac{\partial f_*}{\partial K}) \right]=0.
\end{eqnarray}
Therefore, we have $P^2(r)=0$ and $f_* - K \frac{\partial f_*}{\partial K}=0$ for every $r\in [0,\infty)$ which substituted back in 
(\ref{DI}) give $H^2(r)=0$ for $r\in [0,\infty)$. As in the previous case we can conclude that the map $\varphi$ is a constant map $\varphi=\varphi_0$ with $f_*(\varphi_0,K=0)=0$. Then equations (\ref{DRE})-(\ref{EQFF}) reduce again to the vacuum, static and spherically symmetric Einstein equations whose unique asymptotically flat solution with a regular center is the flat metric.  

Lets  us mention that our theorems proven above in the particular case of canonical scalar fields recover the results of \cite{Heusler_1992}, however the method of proving is completely different from the method of \cite{Heusler_1992}.

The last no-hair theorem is related to the so-called reflecting compact objects (stars) \cite{Hod_2016},\cite{Peng_2018}. More precisely we would like to answer the following question:  Can regular compact, spherically symmetric  reflecting   objects 
(which possess no event horizons) support nontrivial, self-gravitating  non-canonical (and canonical) static multi scalar field configurations in their exterior spacetime regions?  A  rigorous answer to this question  can be given only for Neumann boundary conditions. In other words we consider  a reflecting compact object with a coordinate radius ${\cal R}$ at which we have $\partial_r\varphi^a|_{\cal R}=0$.

\medskip
\noindent

{\bf Theorem 3} {\it Let us  consider self-gravitating non-canonical multi-scalar map $\varphi: (M,g)\to ({\cal E}_N, \gamma)$ with the action 
	(\ref{action}) and let the function $f_{*}(\varphi, K)$ satisfy the inequality $f_* - K \frac{\partial f_*}{\partial K}\ge 0$. 
	Then the static, asymptotically flat,  spherically symmetric reflecting compact objects with Neumann boundary conditions   can not support a  non-trivial   self-gravitating non-canonical (and canonical) multi-scalar map $\varphi: (M,g)\to ({\cal E}_N, \gamma)$ in their exterior spacetime regions.}
	
\medskip
\noindent

The proof of the theorem is the same as in the case of the theorem for black holes with the only difference that the integration is from
the surface $r={\cal R}$ of the reflecting object to  infinity.

\section*{Acknowledgements}
DD acknowledges financial support via an Emmy Noether Research Group funded by the German Research Foundation (DFG) under grant
no. DO 1771/1-1. DD is indebted to the Baden-Wurttemberg Stiftung for the financial support of this research project by the Eliteprogramme for Postdocs.  SY would like to thank the University of Tuebingen for the financial support. The partial support by the Bulgarian NSF Grant KP-06-H28/7 and the  Networking support by the COST Actions  CA16104 and CA16214 are also gratefully acknowledged.


\end{document}